\documentclass[prl,reprint,longbibliography] {revtex4-1}

\usepackage{amsmath}

\usepackage[utf8]{inputenc}
\DeclareUnicodeCharacter{03B4}{$\delta$}
\DeclareUnicodeCharacter{2009}{\,}
\DeclareUnicodeCharacter{03BC}{\mbox{$\mu$}}

\usepackage{times}

\usepackage[english]{babel}

\makeatletter
\def\bbl@set@language#1{%
	\edef\languagename{%
		\ifnum\escapechar=\expandafter`\string#1\@empty
		\else\string#1\@empty\fi}%
	\@ifundefined{babel@language@alias@\languagename}{}{%
		\edef\languagename{\@nameuse{babel@language@alias@\languagename}}%
	}%
	\select@language{\languagename}%
	\expandafter\ifx\csname date\languagename\endcsname\relax\else
	\if@filesw
	\protected@write\@auxout{}{\string\select@language{\languagename}}%
	\bbl@for\bbl@tempa\BabelContentsFiles{%
		\addtocontents{\bbl@tempa}{\xstring\select@language{\languagename}}}%
	\bbl@usehooks{write}{}%
	\fi
	\fi}
\newcommand{\DeclareLanguageAlias}[2]{%
	\global\@namedef{babel@language@alias@#1}{#2}%
}
\makeatother

\DeclareLanguageAlias{en}{english}

\usepackage{times,graphicx}
\usepackage[colorlinks=true,linkcolor=blue,urlcolor=blue,citecolor=blue]{hyperref}

\newcommand{\be}{\begin{equation}}
\newcommand{\ee}{\end{equation}}
\newcommand{\bmul}{\begin{multline}}
\newcommand{\emul}{\end{multline}}
\newcommand{\bea}{\begin{eqnarray}}
\newcommand{\eea}{\end{eqnarray}}

\newcommand{\ket}[1]{|#1\rangle}

\newcommand{\dd}{\mathrm{d}}
\newcommand{\eee}{\mathrm{e}}

\newcommand{\bb}[1]{\left( #1 \right)}

\newcommand{\bbcro}[1]{\left[ #1 \right]}
\newcommand{\bbcror}[1]{\left. #1 \right]}
\newcommand{\bbcrol}[1]{\left[ #1 \right.}

\begin{document}
\title{Damping of elementary excitations in one-dimensional dipolar Bose gases}

\author{Hadrien Kurkjian}
\date{\today}
\affiliation{TQC, Universiteit Antwerpen, Universiteitsplein 1, B-2610 Antwerpen, Belgi\"e}
\author{Zoran Ristivojevic}
\affiliation{Laboratoire de Physique Th\'eorique, Universit\'e de Toulouse, CNRS, UPS, 31062 Toulouse, France}

\begin{abstract}

In the presence of dipolar interactions the excitation spectrum of a Bose gas can acquire a local minimum. The corresponding quasiparticles are known as rotons. They are gaped and do not decay at zero temperature. Here we study the decay of rotons in one-dimensional Bose gases at low temperatures. It predominantly occurs due to the backscattering of thermal phonons on rotons. The resulting rate scales with the third power of temperature and is inversely proportional to the sixth power of the roton gap near the solidification phase transition. The hydrodynamic approach used here enables us to find the decay rate for quasiparticles at practically any momenta, with minimal assumptions on the exact form of the interparticle interactions. Our results are an essential prerequisite for the description of all the dissipative phenomena in dipolar gases and have direct experimental relevance.
\end{abstract}
%\date\today
\maketitle

\textit{Introduction.---} At low pressures and temperatures, helium-4 is a remarkable quantum liquid that is superfluid. Landau characterized the latter state by a dissipationless flow of macroscopic objects at small velocities \cite{lifshitz2013statistical}. Another particular feature of the superfluid helium is seen in its spectrum of elementary excitations. While at lowest momenta it is linear, the spectrum possesses a local minimum. The corresponding quasiparticles are known as rotons and have the wavelengths that practically coincide with the mean interparticle distance. Since the  interaction between helium atoms is strong, the roton can be visualized as yet undeveloped Goldstone mode due to an instability toward the crystallization \cite{nozieres_is_2004}. However, such so-called supersolid state that unifies superfluidity with crystalline order has not been so far observed in helium, despite some controversies \cite{kim_probable_2004,day_low-temperature_2007,kim_absence_2012}. 

Another system that shows some similarities with superfluid helium are dipolar Bose 
gases. They can be realized with atoms possessing large dipolar moments, such as chromium, dysprosium, and erbium. Bose-Einstein condensates of those atoms 
are realized \cite{griesmaier_bose-einstein_2005,beaufils_all-optical_2008,lu_strongly_2011,aikawa_bose-einstein_2012}, 
which opened new avenues for studying various phenomena that originate from the dipolar interaction \cite{lahaye_physics_2009,baranov_condensed_2012}. Some of them are the striped states \cite{wenzel_striped_2017}, the quantum droplets \cite{ferrier-barbut_observation_2016,chomaz_quantum-fluctuation-driven_2016}, and the elusive supersolid state \cite{bottcher_transient_2019,chomaz_long-lived_2019,tanzi_observation_2019,natale_excitation_2019,guo_low-energy_2019,tanzi_supersolid_2019}.

Trapped dipolar Bose gases can exhibit a quasiparticle spectrum with a 
roton minimum \cite{odell_rotons_2003,santos_roton-maxon_2003}. 
This occurs because the dipolar interaction cannot be described only 
by a short-range pseudopotential, but it must also include an anisotropic
long-range part, in order to correctly describe the low-energy scattering 
between bosons \cite{baranov_condensed_2012}. The quasiparticle 
spectrum in weakly-interacting Bose gases is determined by the Bogoliubov 
theory \cite{bogolubov_theory_1947} and depends on the Fourier transform 
of the pseudopotential. Since it is described by the two parameters, one for 
the short-range and the other the long-range part, when they are properly tuned, 
the local minimum can develop in the spectrum. A recent experiment 
\cite{chomaz_observation_2018} have confirmed the presence of rotons in the dipolar Bose gas.

The current understanding of the properties and the dynamics of dipolar gases
is limited due to the lack of dissipative mechanisms in most theoretical 
descriptions. Without dissipation, one can neither describe
the post-quench relaxation observed in Ref.~\cite{chomaz_observation_2018},
nor predict the thermodynamic quantities related to ergodicity, such as the gas viscosity
\cite{Khalatnikov1949,Zwerger2011} and loss of phase coherence \cite{ilzhofer_phase_2019,brouillther}.
In a well-isolated gas, dissipation arises primarily from the interactions
between the quasiparticles, which allow the system to reach equilibrium.
A prerequisite to understand the dissipative dynamics
is thus to compute the quasiparticle lifetime, which is the purpose of this work. 

At zero temperature, rotons and all quasiparticle excitations at lower momenta in a dipolar Bose gas are stable. Being slower than the sound velocity, those quasiparticles cannot emit phonons due to the conservation laws of momentum and energy. Such scenario resembles to the absence of Cherenkov radiation at small velocities. However, the quasiparticles do decay at finite temperature. The dominant process for the damping
of a subsonic quasiparticle involves its scattering with another thermally excited quasiparticle, where two new quasiparticles become created 
\cite{Khalatnikov1949,gangardt_quantum_2010,Andreev2012,landaukhalat,amorbf,Penco2018}. The resulting rate typically scales as a power-law of temperature. 

In this paper we study the damping of energetic quasiparticles (including rotons) in a one-dimensional dipolar Bose gas. This process is controlled by the backscattering of thermal phonons. We find the low-temperature rate that scales as the third power of temperature, $T^3$. The hydrodynamic approach employed here to describe the interaction between phonons and energetic quasiparticles is not limited to the weak-coupling
regime, and does not require a specific form of the interaction potential. However, in order to describe roton damping, we must require specific interactions that lead to the formation of a roton minimum. Near the solidification phase transition, when the roton
gap $\Delta$ is much smaller than the other relevant energy scales (except temperature), we find a rate diverging as $1/\Delta^{6}$. Our results pave the way to a description of the post-quench relaxation dynamics of a dipolar Bose gas \cite{pretherm}.

\begin{figure}
\includegraphics[width=\columnwidth]{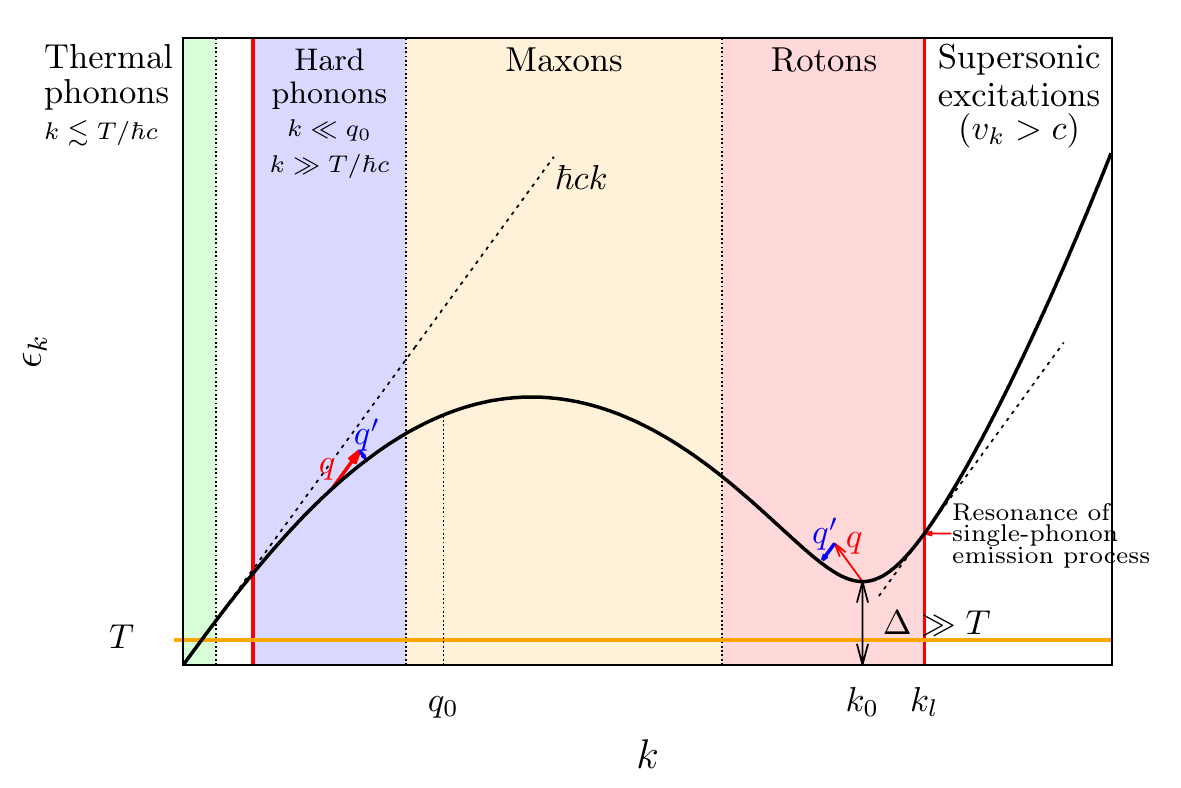}
\caption{\label{fig:schema} Rotonic excitation spectrum of a one-dimensional dipolar Bose gas obtained from the Bogoliubov theory \cite{Santos2007}. 
The damping of energetic excitations of the energy $\epsilon_k\gg T$, i.e., hard phonons, maxons, and rotons is controlled by the backscattering off thermal phonons. Our theory is valid until quasiparticles become supersonic with velocities $v_k>c$ (right vertical red line).}
\end{figure}

\textit{Damping of energetic quasiparticles.---} We consider a gas of bosons with a dipolar interaction between particles in a (quasi-)one-dimensional geometry (studied theoretically in Ref.~\cite{Santos2007} and recently realized
experimentally \cite{Ferlaino2018}). We assume that the gas is prepared at low temperature, $T\ll \Delta,mc^2$, where $mc^2$ denotes the characteristic energy scale for phonons. By $m$ is denoted the mass of particles, $c$ is the sound velocity, while the Boltzmann constant is set to unity. Subsonic quasiparticles with the dispersion $\epsilon_k$ are characterized by the velocity $v_k=\dd\epsilon_k/\hbar\dd k$ that is smaller than $c$. They can decay only due to scattering off thermally excited quasiparticles. At very high momenta, the quasiparticles become supersonic, see Fig.~\ref{fig:schema}. In this case they can decay already at $T=0$ by emitting phonons, which is not precluded by the conservation laws \cite{tan_relaxation_2010,ristivojevic_decay_2016}.

The collisionless damping rate \footnote{The thermal phonons of the characteristic wavevector $q_{\mathrm{th}}=T/\hbar c$ have a lifetime that diverges when $T\to 0$. Therefore, $\epsilon_k\gg \hbar\Gamma_{q_{\mathrm{th}}}$, which justifies the collisionless approximation.} associated to a subsonic quasiparticle of the energy $\epsilon_{k}\gg T$ (see Fig.~\ref{fig:schema}) can be computed using the Fermi golden rule:
\begin{align}
\Gamma_k=\frac{2\pi}{\hbar}\sum_{q,q',k'} \frac{|\mathcal{A}_{if}|^2}{L^2}\delta(\epsilon_k+\hbar\omega_q-\epsilon_{k’}-\hbar\omega_{q'})n_{q}(1+n_{q'}).
\label{Gammakdef}
\end{align}
Here $\omega_q=c|q|$ is the phonon frequency, $L$ is the system size, while $n_{q}=1/[\exp(\hbar\omega_q/T)-1]$ 
denotes the Bose occupation factor~\footnote{The thermal population of energetic quasiparticle modes is negligible in the studied low-temperature case.}. 
In Eq.~(\ref{Gammakdef}), $\mathcal{A}_{if}$ is the transition amplitude from the initial state, $\ket{i}=\hat\gamma_{k}^\dagger \hat b_{q}^\dagger \ket{0}$ 
with an excitation present in the mode $k$,
to the final state $\ket{f}=\hat\gamma_{k'}^\dagger \hat b_{q'}^\dagger \ket{0}$.
By $\hat b^\dagger$ and $\hat \gamma^\dagger$ are denoted the bosonic creation operators for the phonon and the energetic quasiparticle, respectively. The delta function in Eq.~(\ref{Gammakdef}) accounts for the energy conservation.

The transition amplitude $\mathcal{A}_{if}$
can be computed with minimal assumptions, in particular not assuming that the gas is weakly interacting, using quantum hydrodynamics \cite{Khalatnikov1949,Penco2018,Andreev2012,amorbf}.
Within this theory, it is sufficient to consider the cubic residual interaction among phonons given by the Hamiltonian 
\begin{align}
\hat{H}_{\rm ph}=\sum_q \hbar\omega_q \hat b_q^\dagger \hat b_q +\sum_{q,q'} \frac{\mathcal A_{\rm 3}(q,q')}{\sqrt{L}} \bb{\hat b_{q+q'}^\dagger \hat b_q \hat b_{q'}+\textrm{H.c.}}.
\label{Hph}
\end{align}
The matrix elements $\mathcal A_{\rm 3}$ follows from the hydrodynamic equations of motion \cite{Khalatnikov1949,annalen} and have the form
\begin{align}
\mathcal{A}_{\rm 3}(q,q')={}&\frac{mc^2}{\sqrt{32\rho}}\sqrt{\frac{ q q' (q+q')}{ q_0^3}}\biggl[\frac{\rho^2}{c^2}\frac{d}{d\rho}\left(\frac{c^2}{\rho}\right)+\textrm{sgn}(qq') \notag\\
&+\textrm{sgn}(q(q+q'))+\textrm{sgn}(q'(q+q')) \biggr],
\end{align}
where $\rho$ is the (mean) fluid density and $q_0=mc/\hbar$. The hydrodynamics describes
the energetic quasiparticles perturbed by
the phonon field within a local density approximation as
\begin{equation}
\hat H_{\rm qp}=\frac{1}{2}\left[{\epsilon\boldsymbol{(}\hat p,\rho+\delta\hat\rho(\hat r)\boldsymbol{)} +\hat p \hat v(\hat r)}+\mbox{H.c.}\right].
\label{H}
\end{equation}
Here $\epsilon(\hat p,\rho)$ is the Hamiltonian of the unperturbed quasiparticle in first quantization, where
$\hat r$ is its position and $\hat p$ its momentum operator.
By $\delta\hat\rho$ and $\hat v$ are denoted, respectively, the density and the superfluid velocity perturbations caused by the phonons. Expanded to a quadratic order at small $\delta\hat\rho\ll\rho $ and expressed in second quantization, Eq.~(\ref{H}) becomes
\begin{align}
\hat{H}_{\rm qp}={}&\sum_{k} \epsilon_k \hat \gamma_{k}^\dagger \hat \gamma_{k} + \sum_{k,q} \frac{\mathcal{A}_1(k,q)}{\sqrt{L}} \bb{\hat \gamma_{k+q}^\dagger \hat \gamma_{k} \hat b_q + \textrm{H.c.}}\notag\\ 
&+\sum_{k,q,q'} \frac{\mathcal{A}_2(k,q,q')}{L} \hat \gamma_{k+q-q'}^\dagger \hat b_{q'}^\dagger \hat \gamma_{k} \hat b_q.
\end{align}
The phonons and the energetic quasiparticle are coupled by
\begin{gather}
\label{A3}
\mathcal{A}_1(k,q)=\sqrt{\frac{ \rho |q|}{2q_0}} \frac{\partial_\rho \epsilon_k + \partial_\rho \epsilon_{k+q}}{2} 
+\sqrt{\frac{\hbar c}{2m\rho|q|}} q\bb{k+\frac{q}{2}},\notag\\ 
\label{A4}
\mathcal{A}_2(k,q,q')=\frac{\rho \sqrt{|qq’|}}{2q_0} \frac{\partial^2_\rho\epsilon_k + \partial^2_\rho\epsilon_{k+q-q'}}{2}.
\end{gather}
Note the symmetry towards the exchange $k\leftrightarrow k'=k+q$ in $\mathcal{A}_1$
and $k\leftrightarrow k'=k+q-q'$ in $\mathcal{A}_2$, which is a consequence of the hermiticity of the Hamiltonian (\ref{H}).
Computing the transition amplitude in
second-order perturbation theory 
(not forgetting the contribution of three-phonon residual interaction \cite{petkovic_dynamics_2016,erramorbf}), on the mass shell we obtain
\begin{align}
\label{A2eff}
&\mathcal{A}_{if}=\biggl[\mathcal{A}_{2}(k,q,q')+\frac{\mathcal{A}_1(k-q',q') \mathcal{A}_1(k-q',q)}{\epsilon_{k’}-\hbar\omega_{q}-\epsilon_{k'-q}}\notag\\
&
+\frac{\mathcal{A}_1(k,q) \mathcal{A}_1(k',q')}{\hbar\omega_q + \epsilon_k-\epsilon_{k+q}} 
+\frac{2\mathcal{A}_{\rm 3}(q',q-q') \mathcal{A}_1(k,q-q')}{\hbar\omega_{q}-\hbar\omega_{q-q'}-\hbar\omega_{q'}}\notag\\
&+\frac{2\mathcal{A}_{\rm 3 }(q,q'-q) \mathcal{A}_1(k',q'-q)}{\epsilon_{k}-\epsilon_{k'}-\hbar\omega_{q-q'}} \biggr]\delta_{k+q,k’+q’}.
\end{align}
To compute the damping rate \eqref{Gammakdef} at low temperature,
we need the amplitude (\ref{A2eff}) expressed at the leading order in small $q$ and $q'$. The energy conservation constraint of Eq.~(\ref{Gammakdef}) 
has non-trivial solutions only for the backscattering events, i.e., at  $q q'<0$. It leads to the relation 
\begin{equation}
q'=-q\times\begin{cases}
(c-v_k)/(c+v_k) +O(q), \quad &kq>0, \\
(c+v_k)/(c-v_k)+O(q), \quad &\ kq<0.
\end{cases}
\end{equation}
Since the quasiparticle scattering off a phonon experiences a small energy change, we expand it
to second order around the initial energy $\epsilon_k$ as
\begin{equation}
\epsilon_{k'}(\rho)=\epsilon_{k}(\rho)+\hbar v_{k}(\rho)(k'-k)+\frac{\hbar^2(k'-k)^2}{2m^*(k,\rho)}+O{(k'-k)^3}.
\label{epsk}
\end{equation}
In the limit $q\to0$, the on-shell amplitude takes the form $\mathcal{A}_{if}={\hbar c \sqrt{|q q'|}}{ Y}_k \delta_{k+q,k'+q'}/{2\rho}$, where
\begin{widetext}
\begin{equation}
{ Y}_k = \frac{\rho^2}{mc^2(c^2-v_k^2)} \bbcro{(c^2-v_k^2)\partial^2_\rho \epsilon_k-(\partial_\rho\epsilon_k)\partial_\rho(c^2-v_k^2)+\frac{\hbar^2c^2k^2}{m^*\rho^2}-\frac{(\partial_\rho\epsilon_k)^2}{m^*}-\frac{2\hbar ck}{\rho}\bb{v_k\partial_\rho c - c\partial_\rho v_k}}, \label{Aif}
\end{equation}
\end{widetext}
in agreement with Ref.~\cite{Andreev2012}.
Equation (\ref{Aif}) has a divergence when $v_k$ approaches $c$. At high momenta this occur after the roton minimum when the quasiparticle  becomes supersonic, see Fig.~\ref{fig:schema}. Such singularity physically denotes the possibility for the decay by a single phonon emission. The same type of divergence in $Y_k$ also exists when the quasiparticle approaches the phonon regime. Those two thresholds are the boundaries of validity of our hydrodynamic approach (see the vertical red lines in Fig.~\ref{fig:schema}).

Using the calculated on-shell amplitude we can now evaluate the damping rate (\ref{Gammakdef}). To the leading order in temperature we obtain
\begin{align}
\Gamma_k\underset{T\to0}{\sim}\frac{T^3  Y_k^2}{8\pi\hbar^3 c^2\rho^2} \left[\frac{c\,J\left(\frac{c-v_k}{c+v_k}\right)}{c+v_k} +\frac{c\,J\left(\frac{c+v_k}{c-v_k}\right)}{c-v_k}\right], 
\label{Gammak}
\end{align}
where $Y_k$ is given by Eq.~(\ref{Aif}), while the dimensionless  
$J(x)=x\int_0^{+\infty} \dd\tilde{q}\tilde{q}^2  [1+\tilde n(x\tilde{q})]\tilde n(\tilde{q})$, with $\tilde n(\tilde q)=1/(e^{\tilde q}-1)$. The rate (\ref{Gammak}) is our main result. It shows that the low-temperature quasiparticle decay rate universally scales with the third power of temperature. The expression (\ref{Gammak}) is general and independent of the particle interaction strength. The $k$-dependent part of the rate is contained in $\Gamma_k$ and depends on the specific details of the spectrum. We study now Eq.~(\ref{Gammak}) in more details.

\textit{Hard phonon regime.---} Since $T\ll mc^2$, there exists a regime of ``hard phonons’’ with the characteristic energy $T\ll \epsilon_k\ll mc^2$, such that our hydrodynamic approach applies. We consider the general form of the spectrum of the energetic quasiparticle, i.e., a hard phonon, 
\begin{equation}
\epsilon_k=\hbar c(\rho)|k|\bb{1+\frac{\gamma(\rho)}{8} \frac{k^2}{q_0^2} }+ O\bb{{ k}/{q_0}}^4.
\label{qpec}
\end{equation}
Here the coefficient of the small correction term is negative, $\gamma(\rho)<0$, which ensures that the excitation branch is subsonic (i.e., concave) and that the zero-temperature decay by phonon emission  \cite{Matveev2016} does not occur. Expanding Eq.~\eqref{Aif} at small $k$ we obtain $Y_k=-\mathcal{A}k/q_0+O( k/q_0)^2$, where
\begin{align}
\mathcal{A}=1-2\rho \frac{c'}{c}-\rho^2 \frac{c'^2}{c^2}-\rho^2\frac{c''}{c}+\rho\frac{\gamma'}{\gamma} \left(1+\rho\frac{c'}{c}\right).
\end{align}
By prime we denote the derivative with respect to $\rho$. Using $J(x)\sim 2\zeta(3) x$ at $x\to\infty$,
the rate (\ref{Gammak}) becomes
\begin{equation}
\Gamma_k\underset{\substack{T\to0 \\  k/q_0\to 0}}{\sim} \frac{32\zeta(3)}{9\pi} \mathcal{A}^2\frac{q_0^2}{ \gamma^2k^2} \frac{T^3}{\hbar^3 c^2\rho^2}. 
\label{pttk}
\end{equation}
Equation (\ref{pttk}) becomes very large at small $k$, which reflects  the denominator in Eq.~\eqref{Aif}. 
Our approach requires the quasiparticle energy correction $\delta\epsilon_k=\epsilon_k-\hbar c|k|$ of Eq.~(\ref{qpec}) 
not to be thermally smeared, i.e., $\delta\epsilon_k\gg T$. Therefore the rate (\ref{pttk}) applies for phonons 
of wavevectors satisfying $q_{\rm th}^{1/3}q_0^{2/3}\ll |k|\ll q_0$ with the thermal wavevector $q_{\rm th}=T/\hbar c$. 
To study the regime of $k$ comparable to $q_{\rm th}^{1/3}q_0^{2/3}$, our phonon Hamiltonian \eqref{Hph} should be further expanded to describe four-phonon interactions \cite{landaukhalat},
while the phonon dispersion should be now taken with the correction, as in Eq.~\eqref{qpec}.
In fact, this regime has been studied microscopically \cite{Pustilnik2013,ristivojevic_decay_2014}.
For $k=\alpha q_{\rm th}^{1/3}q_0^{2/3}$ and $\alpha$ independent of temperature, 
the $T^3$ power law for $\Gamma_k$ breaks down. However, quite remarkably, one recovers
Eq.~\eqref{pttk} (specified to the Bogoliubov dispersion, see below) 
by taking the limit $\alpha\to\infty$ (see Eq.~(8) in Ref.~\cite{Pustilnik2013}).
This excludes the existence of an intermediate scaling law between 
the regimes $k\approx q_0$ (of which Eq.~\eqref{pttk} gives the low-$k$ limit)
and $k\approx q_{\rm th}^{1/3}q_0^{2/3}$ (of which Eq.~(8) in Ref.~\cite{Pustilnik2013} gives the high-$k$ limit).

\textit{Case of the rotons.---} Rotons are a special case of the energetic quasiparticles considered above, which have a purely quadratic dispersion around the roton minimum $k_0$:
\begin{equation}
\epsilon_{k'}(\rho)=\Delta(\rho)+\frac{\hbar^2 [k'-k_0(\rho)]^2}{2m^*(\rho)}+O(k'-k_0)^3.
\label{epsrot}
\end{equation}
The effective mass $m^*$ is positive for rotons,
but the present discussion also applies to maxons, which have a similar quadratic dispersion but a negative mass, $m^*<0$. For such quadratic dispersion, using $J(1)=\pi^2/3$, the damping rate (\ref{Gammak}) acquires a simpler form
\begin{equation}
\Gamma_{k_0}\underset{T\to0}{\sim} \frac{\pi}{12}Y_{k_0}^2\frac{T^3}{ \hbar^3 c^2\rho^2},
\label{Gammak0}
\end{equation}
where one can use Eq.~(\ref{Aif}) at $k=k_0$ or its alternative version
\begin{align}
 Y_{k_0} ={}& \frac{\rho^2}{mc^2}\bbcrol{\bb{\frac{\dd\mu}{\dd\rho}}^2\frac{\dd^2\Delta}{\dd\mu^2}-\frac{\dd\Delta}{\rho\dd\rho}-\frac{1}{m_*}\bb{\frac{\dd\Delta}{c\dd\rho}}^2}\notag\\
&\bbcror{+\frac{\hbar^2(k_0/\rho- \dd k_0/\dd\rho)^2}{m_*}}.
\label{Ak0}
\end{align}
The correspondence with the general form \eqref{Aif}  is seen using $\Delta=\epsilon_{k_0}$, $\dd\Delta/\dd\rho=\partial_\rho\epsilon_{k_0}$, 
$\dd^2\Delta/\dd\rho^2+{(\hbar\dd k_0/\dd\rho)^2}/{m^*}=\partial_\rho^2\epsilon_{k_0}$, $v_{k_0}=0$, and $\hbar\dd k_0/m^*\dd\rho=-\partial_\rho v_{k}\vert_{k=k_0}$, which follows from partially deriving Eq.~\eqref{epsrot} with respect to $\rho$ and $k'$. The sound velocity $c$ is related to the chemical potential of the gas $\mu$ via the relation $mc^2=\rho\dd\mu/\dd\rho$.

\textit{Weak coupling limit.---} At weak interaction, one can calculate analytically the quasiparticle spectrum \cite{Santos2007,Matveev2016}
which takes the Bogoliubov form,
\begin{equation}
\epsilon_k=\sqrt{E_k\bb{E_k+2\rho g_k}},
\label{epsk}
\end{equation}
where $E_k=\hbar^2k^2/2m$ is the kinetic energy of a free boson. By $g_k$ is denoted the Fourier transform of the effective two particle interaction. Postponing the discussion about its specific form for the following paragraph, we now assume that the spectrum has the characteristic form with the roton minimum, see Fig.~\ref{fig:schema}. Equation (\ref{epsk}) enables us to compute the relevant derivatives with respect to the density:
\begin{gather}
\rho\partial_\rho\epsilon_k=\frac{\epsilon_k^2-E_k^2}{2\epsilon_k}, \qquad  \rho^2\partial_\rho^2\epsilon_k=-\frac{(\epsilon_k^2-E_k^2)^2}{4\epsilon_k^3}, \notag\\ 
\rho \partial_\rho v_k= v_k\frac{\epsilon_k^2+E_k^2}{2\epsilon_k^2}-\frac{\hbar k}{m}\frac{E_k}{\epsilon_k}.
\end{gather}
The roton damping rate (\ref{Gammak0}) now depends only on $\Delta$ and $k_0$,
besides the thermodynamic parameters $T,c$, and $\rho$ which already appears in Eq.~\eqref{Gammak0}.
Since those quantities have been measured experimentally, our expression of $\Gamma_{k_0}$ can be tested
without any assumption on $g_k$. At $\Delta\to0$, such that $T\ll\Delta\ll E_{k_0},mc^2$, we find
\begin{equation}
\Gamma_{k_0}\underset{\substack{T\to0 \\ \Delta\to 0}}{\sim} \frac{\pi}{192}\bb{1+R}^2\frac{ E_{k_0}^8}{(mc^2)^2\Delta^6} \frac{T^3}{\hbar^3 c^2\rho^2},
\label{pttDelta}
\end{equation}
which diverges as $\Delta^{-6}$. In Eq.~(\ref{pttDelta}) we have introduced the dimensionless parameter 
\begin{align}
R=\Delta /m^*c^2.
\end{align}
It has a finite nonzero limit at $\Delta\to 0$. We finally notice that the Bogoliubov spectrum (\ref{epsk}) leads to $\mathcal{A}=3(\gamma-1)/2\gamma$, with $\gamma=1+2\rho mg''_0/\hbar^2$, which should be substituted in Eq.~(\ref{pttk}).
\begin{figure}
\includegraphics[width=\columnwidth]{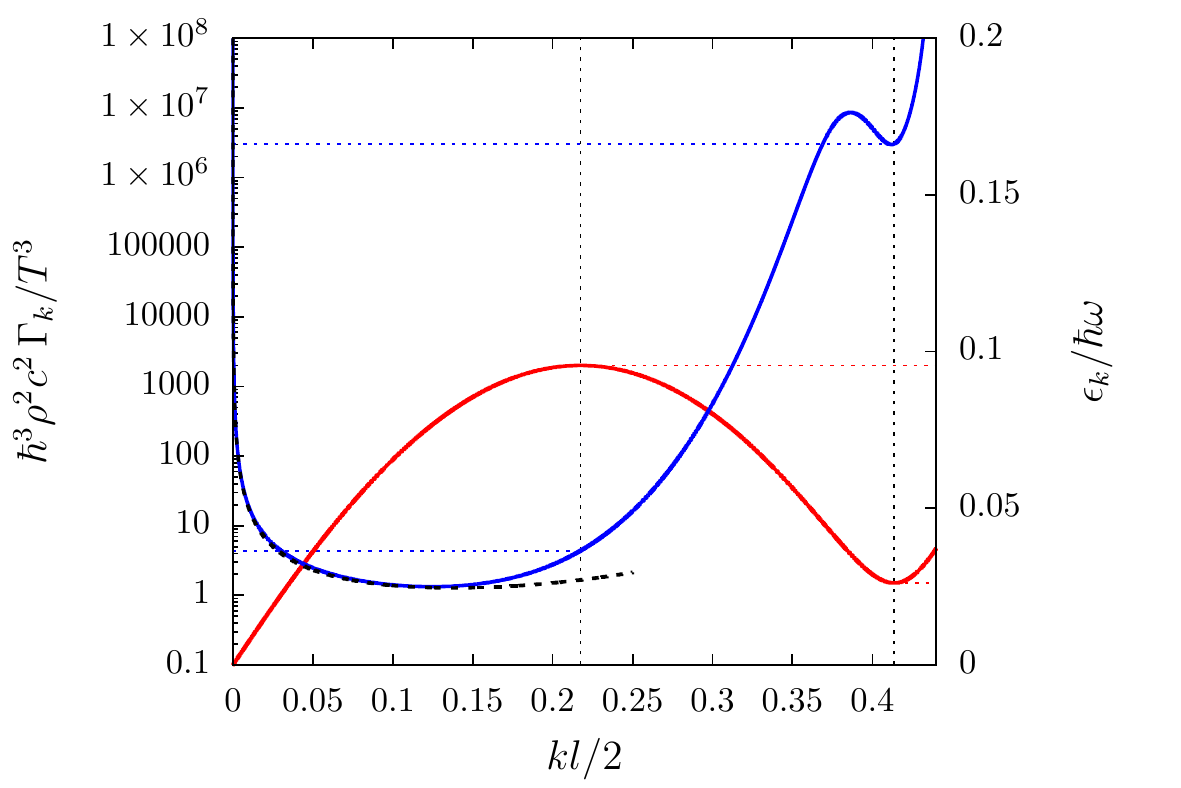}
\caption{\label{fig:enfctdedelta} The quasiparticle spectrum $\epsilon_k$ in units of the transverse trapping energy $\hbar\omega$ (red solid line)
and their rescaled damping rate $\hbar^3\rho^2c^2\Gamma_k/T^3$ (blue solid line) 
in a quasi-one-dimensional dipolar Bose gas as a function of the wavevector $k$ in units of $2/l=2\sqrt{m\omega}/\hbar$.
The dotted vertical black lines show the maxon and roton extrema. The dashed black curve is the hard-phonon asymptote of Eq.~\eqref{pttk}.}
\end{figure}

\textit{Effective interaction potential.---} For the application of our theory on the  realistic model, we consider cylindrically symmetric quasi-one-dimensional Bose gas characterized by the transverse trapping frequency $\omega$ in both transverse directions, and the average dipole moment along the direction of motion~$x$ \cite{Santos2007}. In Fourier space, the effective one-dimensional interaction potential acquires the form
\begin{equation}
g_k=g_{\rm 1D}-\frac{4\alpha d^2}{ l^2}\bbcro{1-\check{k}^2\eee^{\check{k}^2}\Gamma\bb{0,\check{k}^2}},
\label{Z}
\end{equation}
where $g_{\rm 1D}$ is the effective one-dimensional contact coupling constant describing short-range interactions
(including the short-range behavior of the dipolar interactions \cite{Reimann2010,deuretzbacher_erratum_2013}),
$\check{k}=kl/2$ where $l=\hbar/\sqrt{m\omega}$ is the harmonic-oscillator length, $d$ is the dipole moment, while $\Gamma$ denotes the incomplete gamma function. The dipole precesses at high frequency around 
$x$-axis, which leads to an effective dipolar strength $\alpha d^2$ with $-1/2<\alpha<1$ \cite{Pfau2002}. For the interaction potential (\ref{Z}), the sound velocity satisfies $mc^2=\rho (g_{\rm 1D}-V_d)$, where $V_d={4\alpha d^2}/{ l^2}$. The low-energy stability condition requires $g_{\rm 1D}>V_d$. A roton minimum appears at $k_0$ (which solves $\dd \epsilon_k/\dd k\vert_{k=k_0}=0$)
for $g_{\rm 1D}<0$ and $V_{d, {\rm min}}<V_d<V_{d, {\rm max}}<g_{\rm 1D}$. The lower bound $V_{d, {\rm min}}$
is reached at the phase transition $\Delta=\epsilon_{k_0}=0$ and the upper bound $V_{d, {\rm max}}$ when the dispersion
has an inflexion point (when simultaneously $\dd \epsilon_k/\dd k\vert_{k=k_0}=0$ and $\dd^2 \epsilon_k/\dd k^2\vert_{k=k_0}=0$).

\begin{figure}
\includegraphics[width=\columnwidth]{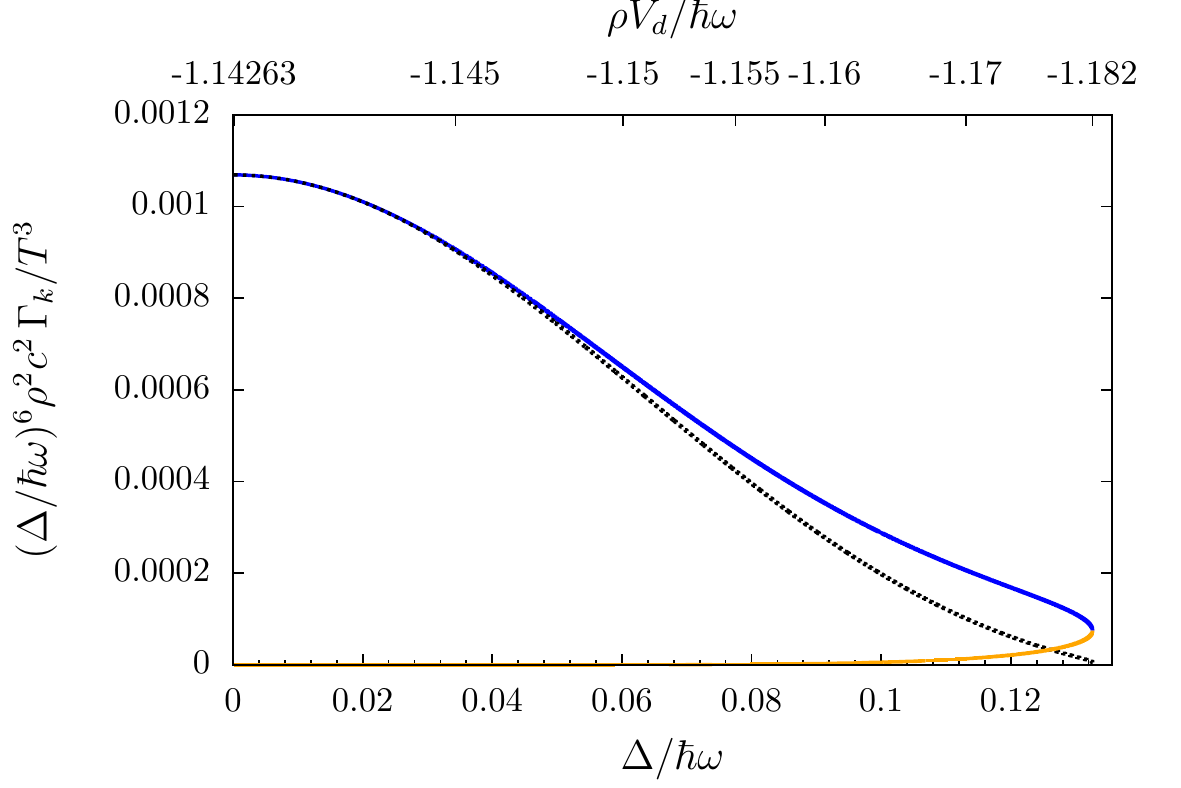}
\caption{\label{fig:enfctdek} Rescaled roton (blue curve) and maxon (orange curve) damping rates $(\Delta/\hbar\omega)^6\rho^2c^2\Gamma_{k}/T^3$ 
as functions of the roton gap (lower $x$-axis) or the dipolar strength $\rho V_d$ (upper $x$-axis) at fixed $\rho g_{\rm 1D}=-\hbar\omega$.
The dashed black line is the corresponding $\Delta\to0$ asymptote of the roton damping rate obtained from Eq.~\eqref{pttDelta}.}
\end{figure}

In Fig.~\ref{fig:enfctdedelta} we plot the quasiparticle dispersion and the rescaled damping rate as functions of the wavevector, at the interaction strengths $\rho g_{\rm 1D}=-\hbar\omega$ and $\rho V_d=-1.144\hbar\omega$.
The rate $\Gamma_k$ diverges at low $k$ as predicted by Eq.~\eqref{pttk} as well as at $k_{l}l/2\simeq0.44$
when the excitations become supersonic ($v_{k_{l}}=c$). It varies several order of magnitude
between the maxon and the roton regimes, the latter being far more susceptible to the decay. Figure \ref{fig:enfctdek} shows the roton damping rate
as a function of the dipolar interaction strength $\rho V_d$, or equivalently of the roton gap $\Delta$,
fixing the short-range coupling constant $\rho g_{\rm 1D}=-\hbar\omega$. The roton minimum exists
for $-1.182\hbar\omega\lesssim \rho V_d\lesssim -1.143\hbar\omega$ and the gap varies from $0$ to about $0.13\hbar\omega$.
The rate $\Gamma_k$ is a monotonic decreasing function of $\Delta$ and diverges as $\Delta^{-6}$ when $\Delta\to0$ as Eq.~\eqref{pttDelta} predicts.

\textit{Discussion.---} The $T^3$ behavior of the damping rate is perhaps our most easily testable prediction. This power
law is specific to one-dimensional systems, and it would change to $T^5$ in the two-dimensional case, \footnote{This follows from a simple power-counting
in the two-dimensional equivalent to Eq.~\eqref{Gammak}} and to $T^7$ in three dimensions \cite{Khalatnikov1949,Chernyshev2012,amorbf}. Attempts to observe the $T^7$ behavior of the roton-phonon damping rate in superfluid helium failed to resolve it from the roton-roton damping rate \cite{Chernyshev2012}, which follows an activation law in $\eee^{-\Delta/T}$. In our case, the $T^3$ law is more favorable, and the capacity to tune the 
gap offered by ultracold gases can also be used to engender a large
prefactor (see, e.g., Fig.~\ref{fig:enfctdek}). We note that the regime
$\Delta\gg T$ is largely within the reach of the state of the art experiments since a temperature of $500 \, \textrm{nK}$ was reached in Ref.~\cite{ilzhofer_phase_2019}. 
For a gap of about $10^2 \, \textrm{Hz}$, this gives $\Delta/T\approx 10^{-3}$.
Bragg-spectroscopic measurements capable of detecting the linewidths of the elementary modes have also been demonstrated in Ref.~\cite{Ferlaino2019_3}. Let us finally notice that an attempt to describe the quasiparticle damping rate for a two-dimensional dipolar Bose gas 
is made in Ref.~\cite{Wilson2013} using the Beliaev-like process where a 
quasiparticle decays into two others, which resulted in an exponentially suppressed rate at low temperatures. 
However, our theory adopted to the two-dimensional case would give a universal $T^5$ power law. 

In this paper we studied the damping of quasiparticles of the dipolar Bose gas with rotons, 
which cannot decay at $T=0$ except at large wavevectors. This should be contrasted 
with the weakly-interacting Bose gas with short-range interactions which has a convex 
spectrum and thus the decay occurs already at $T=0$ \cite{ristivojevic_decay_2016}. 
At weak interaction, the two cases have an overlap at very large wavevectors, 
where the damping rate at $T=0$ approaches a constant value \cite{tan_relaxation_2010,ristivojevic_decay_2016}.

\begin{acknowledgements}

This research was supported by the European Union's Horizon 2020 research and innovation program under the Marie Sk\l odowska-Curie grant agreement number 665501.
\end{acknowledgements}

\providecommand*\hyphen{-}

%\bibliography{references,biblio}
\end{document}